# Reversible Transformation between Isolated Skyrmions and Bimerons

Kentaro Ohara,[1,†] Xichao Zhang,[1,†,*] Yinling Chen,[1,†] Satoshi Kato,[1] Jing Xia,[1] Motohiko Ezawa,[2] Oleg A. Tretiakov,[3] Zhipeng Hou,[4] Yan Zhou,[5] Guoping Zhao,[6] Jinbo Yang,[7] and Xiaoxi Liu[1,*]

[1] *Department of Electrical and Computer Engineering, Shinshu University, 4-17-1 Wakasato, Nagano 380-8553, Japan*

[2] *Department of Applied Physics, The University of Tokyo, 7-3-1 Hongo, Tokyo 113-8656, Japan*

[3] *School of Physics, The University of New South Wales, Sydney 2052, Australia*

[4] *South China Academy of Advanced Optoelectronics, South China Normal University, Guangzhou 510006, China*

[5] *School of Science and Engineering, The Chinese University of Hong Kong, Shenzhen, Guangdong 518172, China*

[6] *College of Physics and Electronic Engineering, Sichuan Normal University, Chengdu 610068, China*

[7] *State Key Laboratory for Mesoscopic Physics, School of Physics, Peking University, Beijing, 100871, China*

(Dated: October 19, 2022)

[†] These authors contributed equally to this work.

[*] Emails:

zhangxichao_jsps@shinshu-u.ac.jp (X.Z.)

liu@cs.shinshu-u.ac.jp (X.L.)

## ABSTRACT

Skyrmions and bimerons are versatile topological spin textures that can be used as information bits for both classical and quantum computing. The transformation between isolated skyrmions and bimerons is an essential operation for computing architecture based on multiple different topological bits. Here we report the creation of isolated skyrmions and their subsequent transformation to bimerons by harnessing the electric current-induced Oersted field and temperature-induced perpendicular magnetic anisotropy variation. The transformation between skyrmions and bimerons is reversible, which is controlled by the current amplitude and scanning direction. Both skyrmions and bimerons can be created in the same system through the skyrmion-bimeron transformation and magnetization switching. Deformed skyrmion bubbles and chiral labyrinth domains are found as non-trivial intermediate transition states. Our results may provide a unique way for building advanced information-processing devices using different types of topological spin textures in the same system.

*Keywords:* *skyrmion, skyrmion bubble, bimeron, topological spin texture, topological transformation, spintronics*

Information bits are indispensable components for any data storage and computing technologies [1]. Spin textures showing non-trivial topology and non-volatility could be functional information bits in future spin-electronic devices[2-17]. The most robust topological spin textures are skyrmions in perpendicularly magnetized systems [2-23] and their bimeron counterparts in in-plane magnetized systems [10-12, 24-31]. Both skyrmions and bimerons can be created and manipulated by external fields and currents [2-31]. Particularly, they have multiple controllable degrees of freedom [27, 32-34], such as the topological charge and helicity, which offer the possibility for non-conventional applications [6, 7, 9-13, 15, 16]and quantum computing [35].

Traditional concept of making a topological spintronic device usually involves only one type of topological spin textures [6, 7, 9-13, 15, 16], either skyrmions or bimerons, which depends on the material properties, such as the perpendicular magnetic anisotropy (PMA). For example, standard skyrmions are formed in perpendicularly magnetized layers with PMA and Dzyaloshinskii-Moriya interaction (DMI) [2-23], while bimerons and deformed skyrmions are formed in in-plane magnetized layers with DMI and in-plane/easy-plane magnetic anisotropy (including crystalline or shape anisotropy) [24-31, 36, 37]. However, it is envisioned that different types of topological spin textures may work jointly in an advanced storage and computing architecture [11, 12], where different topological spin textures may play different roles and form an environment that allows diverse spintronic building blocks to concurrently perform a complex task.

Therefore, it is vital to realize the creation, transformation, and manipulation of skyrmions and bimerons in the same device in a controlled manner. The magnetic-field-induced transformation between skyrmion crystals and bimeron crystals has been demonstrated in experiments [38]. The temperature-induced meron-skyrmion transformation in a single magnetic domain wall has also been realized experimentally [39]. The transformation between vortices and bimerons by applying an in-plane magnetic field has been numerically demonstrated in a Janus-magnet-based multiferroic heterostructure [40]. Moreover, a recent theoretical report suggests that skyrmion tubes can be transformed to bimeron tubes [31]. A most recent work also suggests that a skyrmion tube carrying a higher topological charge can be bifurcated [41]. However, the creation of massive isolated skyrmions and bimerons in a same magnetic layer, as well as their *in-situ* reversible transformation are still elusive and remain to be discovered.

In this work, we show the creation of isolated skyrmions and their subsequent reversible transformation to isolated bimerons in a magnetic disk controlled by synergistic variations of magnetic field and PMA. The synergistic variations of magnetic field and PMA are realized by an omega-shaped microcoil surrounding the magnetic disk, which generates an Oersted field and a thermal flow to the magnetic disk when a constant electric current is injected into the microcoil. The disk is affected by the thermal effect, leading to the decrease of the PMA. This method offers a possibility to build a device hosting both skyrmions and bimerons.

We first experimentally demonstrate the reversible transformation between skyrmions with $Q = \pm 1$ [see Fig. 1(a)] and bimerons with $Q = \pm 1$ [see Fig. 1(a)], where $Q$ being the topological charge defined as $Q = \int \boldsymbol{m} \cdot \left( \frac{\partial \boldsymbol{m}}{\partial x} \times \frac{\partial \boldsymbol{m}}{\partial y} \right) dx dy / 4\pi$ [11, 12]. Figure 1(b) illustrates the experimental device, where a magnetic disk with the diameter of 90 μm on the glass substrate is surrounded by a concentric omega-shaped microcoil in the same x-y plane. The magnetic disk is made of a deposited Pt (0.3 nm)/CoNi (0.7 nm)/Pt (0.5 nm)/CoNi (0.7 nm)/Pt (0.5 nm)/FeCo (0.4 nm)/Pt (1 nm) multilayer [see Fig. 1(c)] with PMA and interfacial DMI. More details about the fabrication process and characterization, including the measurements of saturation magnetization, PMA, DMI, surface morphology, and grain size, are given in the Methods and Supplementary Note 1.

To control the spin textures, we apply an electric current through the omega-shaped microcoil, which generates an Oersted field perpendicular to the disk. We assume that a positive current (i.e., $I > 0$) generates a positive Oersted field pointing at the +z direction, while a negative current (i.e., $I < 0$) generates a negative Oersted field pointing at the -z direction. This assumption holds true as the magnetic disk and microcoil are lying in the same plane. The magnetic disk is initially with a demagnetized state obtained at zero field and room temperature. By slowly scanning the current from +30 mA to -30 mA, and then from -30 mA to +30 mA, we obtained a current-dependent hysteresis loop [see Fig. 1(d)], which is measured by the polar magneto-optical Kerr effect (MOKE) technique (see Supplementary Note 1).

Interestingly it can be seen that the shape of the current-dependent hysteresis loop is different from that of a typical magnetic layer with PMA. When the amplitude of the current $|I|$ is larger than 20 mA, the reduced out-of-plane magnetization of the disk approaches zero with increasing $|I|$. This phenomenon cannot be explained by the current-induced Oersted field, which should lead to the saturation of the magnetization in the out-of-plane directions, i.e., a ferromagnetic state pointing at the +z or -z direction. The MOKE images in Fig. 1(e) corresponding to the current-dependent hysteresis loop show that isolated skyrmions with $Q = -1$ are created in a perpendicularly magnetized background when the current scans from +30 mA to 0 mA, and when the current scans from 0 mA to -20 mA, the skyrmions expand and form chiral labyrinth domains. When the current further scans to -30 mA, the labyrinth domains disappear, and many objects with skyrmion-like contrast are generated in an in-plane magnetized background. Note that the perpendicularly and in-plane magnetized backgrounds can be distinguished from the black-gray-white color scale of the MOKE images, where black and white contrasts indicate out-of-plane magnetization while gray contrast indicates in-plane magnetization. We will confirm that these skyrmion-like objects in the in-plane magnetized disk are so-called bimerons[10-12,24-31] with $Q = \pm 1$. In a similar way, by continuing to scan the current from -30 mA to 0 mA and then to +30 mA, we obtained a reversible transformation process from bimerons to skyrmions with $Q = +1$ and then to bimerons again. Namely, by executing a loop scan of $I = +30$ mA $\rightarrow$ 0 mA $\rightarrow$ -30 mA $\rightarrow$ 0 mA $\rightarrow$ +30 mA (see Supplementary Video 1), we are able to realize the transformation process of bimerons $\rightarrow$ skyrmions ($Q = -1$) $\rightarrow$ bimerons $\rightarrow$ skyrmions ($Q = +1$) $\rightarrow$ bimerons.

As mentioned above, such a transformation cannot be straightforwardly explained as a consequence of the magnetization switching driven by the Oersted field. Considering the fact that the disk exhibits a nearly in-plane magnetized background at large current (i.e., gray contrast), we propose that the PMA is reduced due to the thermal effect at large current. Namely, the magnitude of the PMA should be inversely proportional to the amplitude of the current in the microcoil. Besides, the PMA may also be affected by a thermal stress [42], which could be caused by the thermal expansion of the sample due to the heating effect. In Supplementary Note 1, we show experimental evidences that the variation of the PMA is dominated by thermalization, while the effect of stress is not significant than that induced by the thermal effect.

For the purpose of examining our hypothesis and advancing the understanding of the reversible skyrmion-bimeron transformation, we carried out simulations using the MuMax3 software [43]. The simulated sample is a 1-nm-thick disk with the diameter of 2000 nm, which is surrounded by a concentric and coplanar circular current-carrying coil of diameter 2200 nm. Material parameters and simulation settings are given in Supplementary Note 2. We assume that a positive current injected into the coil generates a positive Oersted field perpendicular to the disk. The direction and magnitude distribution of the Oersted field is calculated numerically using the MATLAB magnetic field simulator [44] based on the Biot-Savart law [45] [see Fig. 2(a)]. On the other hand, we assume that the disk has a current-dependent PMA according to our experimental results (see Fig. 1 and Supplementary Note 1). Namely, the PMA constant $K$ is assumed to be proportional to -$I^2$ with $I$ being the current amplitude, i.e., $K = K_{min} - \left(\frac{K_{max} - K_{min}}{I_{max}^2}\right) \cdot I^2$, as depicted in Fig. 2(b). We have the default values: $K_{max}$ = 0.66 MJ m$^{-3}$, $K_{min}$ = 0.40 MJ m$^{-3}$, and $I_{max}$ = ±30 mA. Such a current-dependent PMA ideally mimics the temperature-induced PMA variation in experiments. Other parameters are assumed to be independent of the current to simplify the model. Indeed, we have computationally investigated the effects of different magnetic parameters (see Supplementary Note 3).

The initial state is relaxed from a randomly distributed spin configuration at zero field, mimicking the demagnetized state in experiments. By increasing the current from 0 mA to +30 mA and then scanning the current between +30 mA and -30 mA with a scan rate of 1 mA per stage, a full hysteresis loop is obtained [see Fig. 2(c)]. For each stage of current change, the system is relaxed by minimizing the total energy in MuMax3 [43], which reproduces the experimental scanning procedure. The hysteresis loop excluding the initial magnetization curve given in Fig. 2(d) shows a good qualitative agreement with the experimental one [see Fig. 1(d)]. Figure 2(e) shows the selected snapshots of the spin textures. It shows that both skyrmions and bimerons are found at certain ranges of the current, as indicated by yellow and cyan shadows in Figs. 2(b) and 2(d). Skyrmions are found in the perpendicularly magnetized background at relatively larger $K$ (i.e., relatively smaller $|I|$ around $|I|$ = 0 mA), while bimerons are found in the in-plane magnetized background at relatively smaller $K$ (i.e., relatively larger $|I|$ near $|I|$ = 30 mA).

We focus on the segment corresponding to the switching during the current scan $I$ = 0 mA → -30 mA, which is highlighted in green in Figs. 2(d) and 2(e). As the current scans from 0 mA to -30 mA, the Oersted field pointing at the -$z$ direction increases while the PMA decreases with the current in a synergistic manner. In principle, the increasing Oersted field initially leads to the magnetization switching from the +$z$ direction to -$z$ direction, while the decreasing PMA further leads to the reorientation of magnetization from the ±$z$ direction to the $x$-$y$ plane due to the effect of demagnetization. The joint effect of the Oersted field and PMA fosters the skyrmion-bimeron transformation, as described follows: **(a)** the Oersted field first leads to the expansion of isolated skyrmions [see Fig. 2(e), $I$ = 0 mA → -5 mA]; **(b)** the skyrmions further expand and get deformed by the Oersted field, which forms island spin textures [see Fig. 2(e), $I$ = -5 mA → -17 mA]; **(c)** the deformed skyrmions near the disk edge are destroyed by touching the edge when the Oersted field is larger than a certain threshold, which leads to some chiral stripe domain walls in the disk and forms labyrinth domain pattern [see Fig. 2(e), $I$ = -17 mA → -20 mA]; **(d)** the effect of reduced

PMA becomes stronger as the current increases, and consequently the magnetization start to reorientate their directions toward the in-plane directions due to the demagnetization [see Fig. 2(e), $I$ = -20 mA → -25 mA]; **(e)** due to the reorientation of magnetization from the out-of-plane directions to the in-plane directions, the labyrinth domain pattern ceases to exist and some bimerons are generated [see Fig. 2(e), $I$ = -25 mA → -26 mA]; **(f)** the magnetization will be dominated by the strong demagnetization effect when the PMA is too small, which annihilates most bimerons and forms an in-plane ferromagnetic state [see Fig. 2(e), $I$ = -26 mA → -30 mA].

Such a transformation from skyrmions to bimerons, as illustrated in Figs. 2(e) and 2(f), is mediated by the formation, deformation, and annihilation of non-trivial chiral spin textures. The reorientation of magnetization from the out-of-plane to in-plane directions plays an important role for the generation of bimerons as bimerons are topological counterparts of skyrmions in in-plane magnetized systems [10-12, 24-31]. The simulated transformation between skyrmions and bimerons is a reversible process [see Fig. 2(e)], namely, by scanning the current in sequence of $I$ = +30 mA → -30 mA → +30 mA (see Supplementary Video 2), one can create skyrmions and bimerons in the order of bimerons → skyrmions ($Q$ = -1) → bimerons → skyrmions ($Q$ = +1) → bimerons, where both skyrmion-bimeron and bimeron-skyrmion transformations are realized. The simulation results are in line with experimental observations given in Fig. 1. In Supplementary Note 3, we also demonstrate the dependencies of the transformation on different magnetic parameters and disorders, which suggests the transformation is robust for a certain range of parameters.

We note that in both experiments and simulations the density of skyrmions with $Q$ = +1 is smaller than that of skyrmions with $Q$ = -1 [see Figs. 1(e) and 2(e)], which is a phenomenon determined by mediated labyrinth domain patterns and may depend on the history of current scan (i.e., the initial scanning direction and maximum current amplitude). In addition, in experiments the bimeron-to-skyrmion transformation is not mediated by obvious labyrinth domains [see Fig. 1(e)], possibly due to the slightly different thermal accumulations or stresses generated by increasing and decreasing $|I|$ procedures, respectively.

In Fig. 3, we focus on the simulated segment of $I$ = 0 mA → -30 mA (indicated by green color in Figs. 2 and 3) and further analyze the skyrmion-bimeron transformation. By scanning the current from 0 mA to -30 mA, the current-dependent magnetization curve indicates that the magnetization reorientation happens soon after the Oersted-field-induced magnetization switching [see Fig. 3(a)]. The total system energy increases with increasing current amplitude. Most interestingly, the total topological charge of the system changes from -19 (i.e., 19 skyrmions with $Q = -1$) to -9 (i.e., nine bimerons with $Q = -1$, a meron with $Q = -0.5$, and the tilted disk edge with $Q = +0.5$) when the current scans from 0 mA to -26 mA [see Figs. 3(b)]. The sign of the topological charge of a skyrmion is determined by the perpendicularly magnetized background. Therefore, as the bimerons are transformed from some of the skyrmions, the sign of the topological charge of the bimerons is identical to that of the skyrmions.

To justify that the skyrmion-like contrasts observed in our experiments are bimerons, in Fig. 4 we compare our experimental MOKE images to that of simulated skyrmions and bimerons. Figures 4(a) and 4(b) show the grayscale image and contour analysis (see Supplementary Note 1) of the skyrmion. The grayscale image and contour analysis of the bimeron are given in Figs. 4(c) and 4(d). The contours of experimentally observed spin textures are in a good qualitative agreement with simulated results. The agreement between experimental and simulation results justifies our supposition that the skyrmion-like objects in the in-plane magnetized background are bimerons. We also experimentally demonstrate that the skyrmion-bimeron transformation cannot happen without the assistance of the reduced PMA induced via the current-carrying microcoil. As shown in Fig. 5(a), we apply a uniform out-of-plane magnetic field $B_z$ to switch the magnetization in the magnetic disk. The out-of-plane hysteresis loop measured by the MOKE technique is given in Fig. 5(b), which indicates the constant PMA property of the magnetic disk. Such a hysteresis loop is in a stark contrast to that obtained by scanning the current through the microcoil [cf. Figs. 1(d) and 5(b)]. During the magnetization switching driven by the uniform magnetic field, we do not observe either skyrmions or bimerons [see Fig. 5(c)]. Hence, we believe that the microcoil-induced synergistic variations of Oersted field and PMA are responsible for the creation of skyrmions and bimerons and their mutual transformation mediated by other states.

In conclusions, we have demonstrated the reversible transformation between isolated skyrmions and bimerons in a magnetic disk surrounded by a current-carrying and omega-shaped microcoil. The current injected into the microcoil generates an Oersted field that switches the magnetization of the magnetic disk in the out-of-plane directions. At the same time, the current injected into the microcoil heats the device and results in the increase of the device temperature. Therefore, a temperature-induced decrease of PMA is assumed to be realized in the magnetic disk. We assume that the PMA decreases with the applied current amplitude in a quadratic manner because that a larger current can lead to stronger thermalization and thus stronger temperature-induced variation of the PMA (see Supplementary Note 1). Our simulation results show that a decrease of PMA could lead to the magnetization reorientation toward the in-plane directions, and thus fosters the generation of bimerons. As the Oersted field and PMA are controlled by the microcoil, the reversible transformation between skyrmions and bimerons can be realized by scanning the current in a reasonable range. The formation of bimerons does not require a fully in-plane magnetized background at extremely large current. Our results demonstrate the possibility that two different types of topological spin textures can be hosted by a same magnetic device, which may provide guidelines for building advanced spintronic applications based on different types of isolated topological spin textures.

## METHODS

***Magnetic microdisk fabrication.*** A commercially mask-less photolithography instrument is used to fabricate a micro disk precisely at the center of the omega-shaped one turn coil. The micro disk is composed of a multilayered stack of Pt (0.3 nm)/CoNi (0.7 nm)/Pt (0.5 nm)/CoNi (0.7 nm)/Pt (0.5 nm)/FeCo (0.4 nm)/Pt (1 nm). All the films are deposited by DC magnetron sputtering under an Ar pressure of 2.0 Pa in the chamber with a base pressure under $8 \times 10^{-5}$ Pa at room temperature. Target-to-substrate distance was 70 mm. For all the photolithography process, we are using a 500 nm thick photoresist (AZ5200). Here we point out that our samples have densely packed nanocrystalline with a typical average grain size around 5 nm, which is much smaller than the average skyrmion size of about 500 nm. We also find that our sample has an interface-induced DMI of about 0.44 mJ m$^{-2}$. The details on the Omega-shaped microcoil fabrication, MOKE measurements, and contour analysis are given in Supplementary Note 1.

## ASSOCIATED CONTENT

### Supporting Information

The Supporting Information is available free of charge at [URL].

Additional experimental methods; Experimental characterizations; Simulation methods and parameters; Simulations of the skyrmion-bimeron transformation at different magnetic parameters; Enlarged views of the skyrmion and bimeron (PDF).

Movie 1: Experimental observation of the reversible transformation between isolated skyrmions and bimerons, corresponding to Fig. 1(e) of the main text. We scan the current from +30 mA to -30 mA and then from -30 mA to +30 mA. (MOV)

Movie 2: Simulated reversible transformation between isolated skyrmions and bimerons, corresponding to Fig. 2(e) of the main text. We first increase the current amplitude from zero to +80 mA and then scan the current between +30 mA and -30 mA. The scan rate is set to be 1 mA per stage in the simulation. (MP4)

Movie 3: The photo-elastic microscope video showing the photoelastic measurement to check the current effect of the omega coil. A sufficient large stress is created if we apply an AC current to the one turn omega coil. The frequency of the AC current is 1 Hz. The maximum current is 35 mA. (AVI)

Movie 4: Simulated reversible transformation between isolated skyrmions and bimerons in a rectangle-shaped film. We first increase the current amplitude from zero to +30 mA and then scan the current between +30 mA and -30 mA. The scan rate is set to be 1 mA per stage in the simulation. Other parameters are the same as that used in Fig. 2 of the main text. (MP4)

Movie 5: Simulated reversible transformation between isolated skyrmions and bimerons in the absence of the Oersted field. All parameters are the same as that used in Fig. 2 of the main text, while only the Oersted field is turned off during the simulation. (MP4)

## AUTHOR INFORMATION

### Corresponding Authors

Xichao Zhang; E-mail: zhangxichao_jsps@shinshu-u.ac.jp

Xiaoxi Liu; E-mail: liu@cs.shinshu-u.ac.jp

**Author Contributions**

X.L. designed and initiated the study. X.L., K.O., Y.C., and S.K. performed the experiments. X.Z. carried out the computational simulations. J.X. calculated the Oersted field. X.Z. and X.L. drafted the manuscript and revised it with input from M.E., O.A.T., Z.H., Y.Z., G.Z. and J.Y. All authors discussed the results and reviewed the manuscript. K.O., X.Z. and Y.C. contributed equally to this work.

**Notes**

The authors declare no competing financial interest.

**Code Availability**

The micromagnetic simulations are performed using the freely available MuMax3 platform, which is publicly accessible at https://mumax.github.io.

**Data Availability**

The data that support the findings of this study are available from the corresponding authors upon reasonable request.

**ACKNOWLEDGEMENTS**

X.Z. was an International Research Fellow of Japan Society for the Promotion of Science (JSPS). X.Z. was supported by JSPS KAKENHI (Grant No. JP20F20363). J.X. was a JSPS International Research Fellow. J.X. was supported by JSPS KAKENHI (Grant No. JP22F22061). M.E. acknowledges support by the Grants-in-Aid for Scientific Research from JSPS KAKENHI (Grant Nos. JP18H03676 and JP17K05490) and the support by CREST, JST (Grant Nos. JPMJCR20T2 and JPMJCR16F1). O.A.T. acknowledges support from the Australian Research Council (Grant No. DP200101027), NCMAS grant, and the Cooperative Research Project Program at the Research Institute of Electrical Communication, Tohoku University. Z.H. acknowledges support by the National Key Research and Development Program of China (Grant No. 2020YFA0309300), the National Natural Science Foundation of China (Grant No. 51901081), and the Science and Technology Program of Guangzhou (Grant No. 202002030052). Y.Z. acknowledges support by Guangdong Special Support Project (Grant No. 2019BT02X030), Shenzhen Fundamental Research Fund (Grant No. JCYJ20210324120213037), Shenzhen Peacock Group Plan (Grant No. KQTD20180413181702403), Pearl River Recruitment Program of Talents (Grant No. 2017GC010293), and National Natural Science Foundation of China (Grant Nos. 11974298 and 61961136006). G.Z. acknowledges support by the National Natural Science Foundation of China (Grant Nos. 51771127, 51571126, and 51772004), and Central Government Funds of Guiding Local Scientific and Technological Development for Sichuan Province (Grant No. 2021ZYD0025). J.Y. acknowledges support by the National Natural Science Foundation of China (Grant No. 51731001). X.L. acknowledges support by the Grants-in-Aid for Scientific Research from JSPS KAKENHI (Grant Nos. JP20F20363, JP21H01364, JP21K18872, and JP22F22061).

## FIGURE CAPTIONS

**Figure 1. Current-controlled reversible transformation between skyrmions and bimerons. a,** Schematic illustrations of isolated skyrmions with $Q = \pm 1$ and bimerons with $Q = \pm 1$. **b,** Schematic illustration of the device structure, where a magnetic disk with the diameter of 90 µm is surrounded by a conducting microcoil in the same $x$-$y$ plane. An electric current flowing through the coil will generate an Oersted field perpendicular to the magnetic disk and alter the magnetic anisotropy due to the current-induced thermal stress. Inset shows a real top-view microscope image of the disk and coil. Scale bar, 50 µm. **c,** Schematic illustration of the magnetic layer structure. The unit of thickness is nm. **d,** Current-dependent hysteresis loop measured by the MOKE technique. The out-of-plane magnetic field is generated by the conducting coil, which depends on the current amplitude $I$. **e,** Sequential MOKE images showing the reversible transformation between isolated skyrmions and bimerons. Deformed skyrmion bubbles and chiral labyrinth domains exist as non-trivial intermediate transition states. The out-of-plane magnetization component is color coded: light gray is into the plane (pointing at $-z$), dark gray is out of the plane (pointing at $+z$). Scale bar, 20 µm.

**Figure 2. Simulation of the reversible transformation between skyrmions and bimerons. a,** Exemplary distribution profiles of the Oersted field along the diameter of the ferromagnetic disk. **b,** The perpendicular magnetic anisotropy constant $K$ as a function of the applied current $I$. Skyrmion and bimeron areas are indicated by yellow and cyan backgrounds, respectively. **c,** Current-dependent hysteresis loop of the out-of-plane magnetization $m_z$, which includes the initial magnetization curve. The segments for $I = 0$ mA $\rightarrow$ +30 mA, +30 mA $\rightarrow$ 0 mA, 0 mA $\rightarrow$ -30 mA, -30 mA $\rightarrow$ 0 mA, and 0 mA $\rightarrow$ +30 mA are colored in red, blue, green, purple, and gold, respectively. **d,** The hysteresis loop excluding the initial magnetization curve, where skyrmion and bimeron areas are indicated by yellow and cyan backgrounds, respectively. **e,** Top-view snapshots of the static spin configurations in the ferromagnetic disk obtained at different current amplitudes. **f,** Zoomed illustrations of selected spin configurations as indicated by green boxes in e (see Supplementary Note 4 for enlarged views of the skyrmion and bimeron structures). The diameter of the skyrmion at $I = -5$ mA is about 96 nm.

**Figure 3. Current-dependent energy and topological charge of the skyrmion-bimeron transformation. a,** Total energy $E_{\text{Total}}$ and out-of-plane magnetization $m_z$ as functions of the current amplitude. The current amplitude decreases from 0 mA to -30 mA, as indicated by the current scanning direction. **b,** Total topological charge of the disk as a function of the current amplitude. Insets show the top-view snapshots of the spin configurations at $I = -5$ mA and $I = -26$ mA. The diameter of the simulated magnetic disk is 2000 nm. Topological charges of isolated spin textures are indicated by numbers.

**Figure 4. Comparison between isolated skyrmions and bimerons. a**, Experimental MOKE image and contour analysis of the skyrmion state obtained at a current amplitude of 20 mA. Scale bar, 20 μm. The out-of-plane magnetization component in MOKE images in indicated in a black-gray-white color bar. **b**, Contour analysis of an isolated skyrmion obtained in computational simulation [see Fig. 3(b) inset]. The out-of-plane magnetization component in contour images is indicated by a rainbow color bar. **c**, Experimental MOKE image and contour analysis of the bimeron state obtained at a current amplitude of -23.5 mA. **d**, Contour analysis of an isolated bimeron obtained in computational simulation [see Fig. 3(b) inset]. The contour analysis images displayed with rainbow color scale highlight both the in-plane and out of-plane spin components of the topological spin textures.

**Figure 5. Magnetization switching driven by a uniform out-of-plane magnetic field. a**, Schematic illustration of the magnetic layer structure. The unit of thickness is nm. The out-of-plane magnetic field is generated by a large wire-wound coil outside the sample holder. **b**, Hysteresis loop measured by the MOKE technique. **c**, Sequential MOKE images showing the magnetization switching process driven by the uniform out-of-plane magnetic field, which is in a stark contrast to that driven by the microcoil-induced variations of Oersted field and in-plane magnetic anisotropy. Scale bar, 20 μm.

# FIGURES

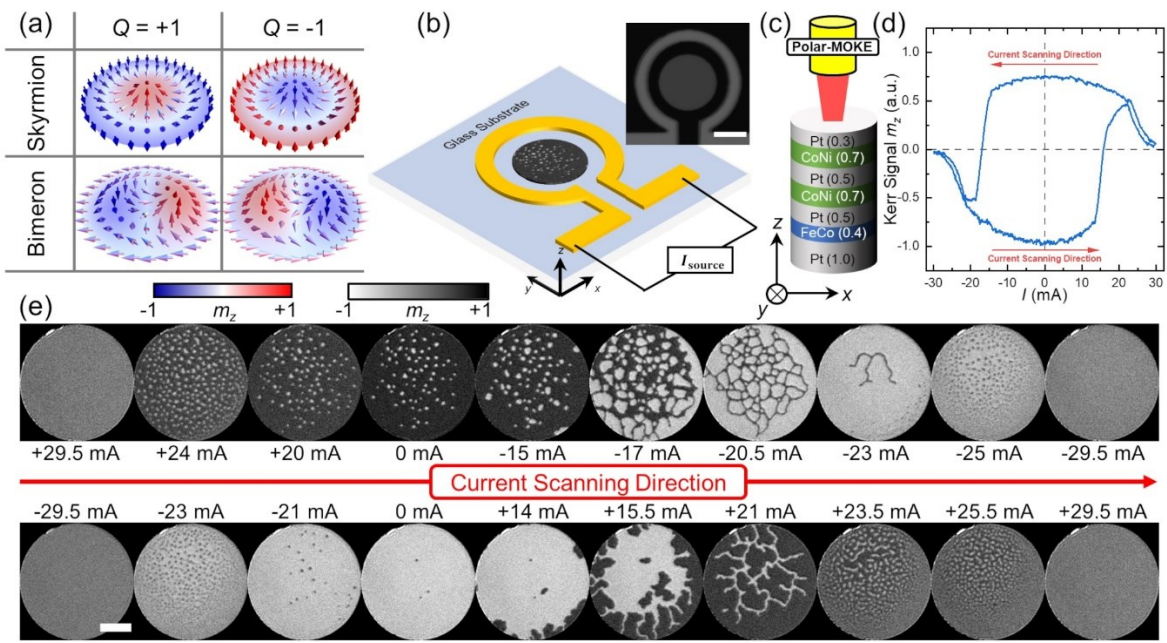

**Figure 1.**

Ohara *et al.*

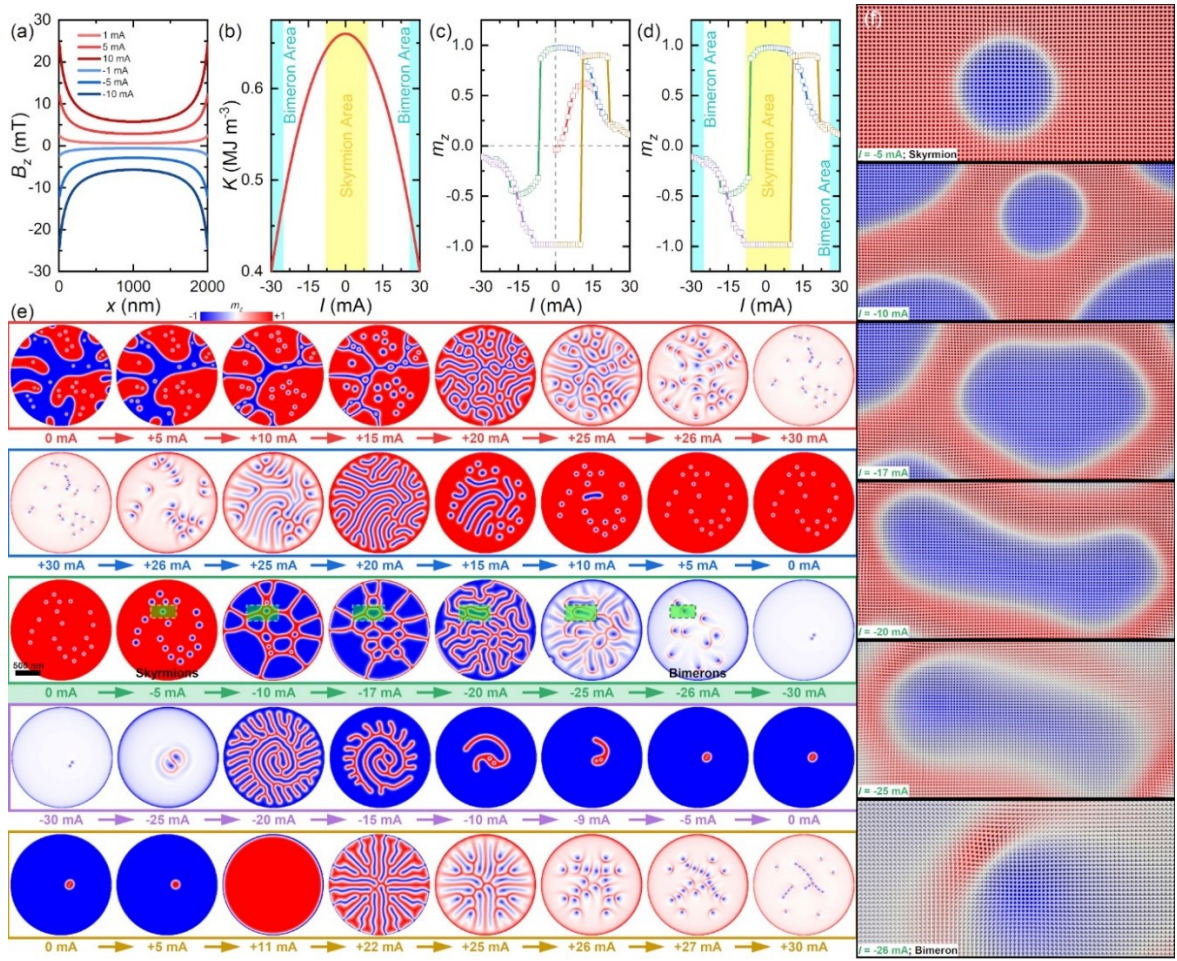

**Figure 2.**

Ohara *et al.*

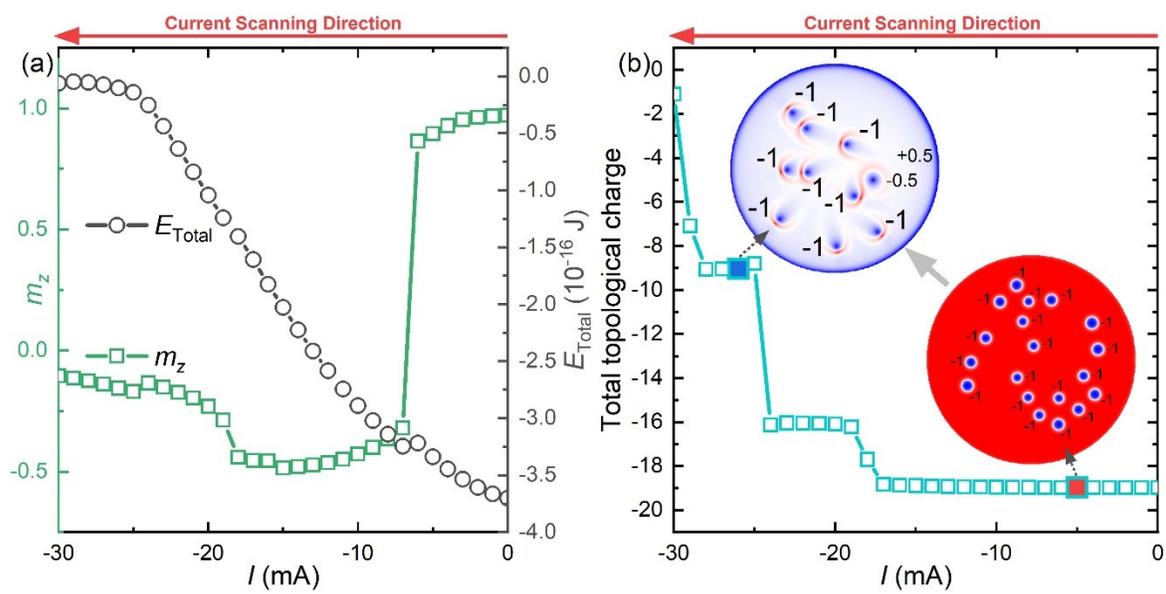

**Figure 3.**

Ohara *et al.*

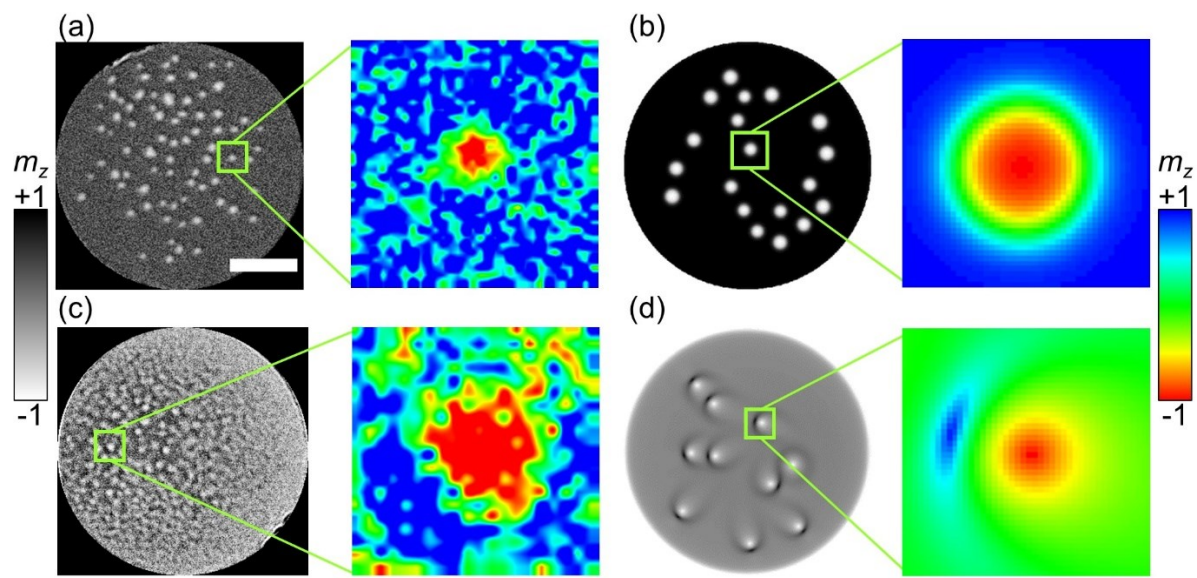

**Figure 4.**

Ohara *et al.*

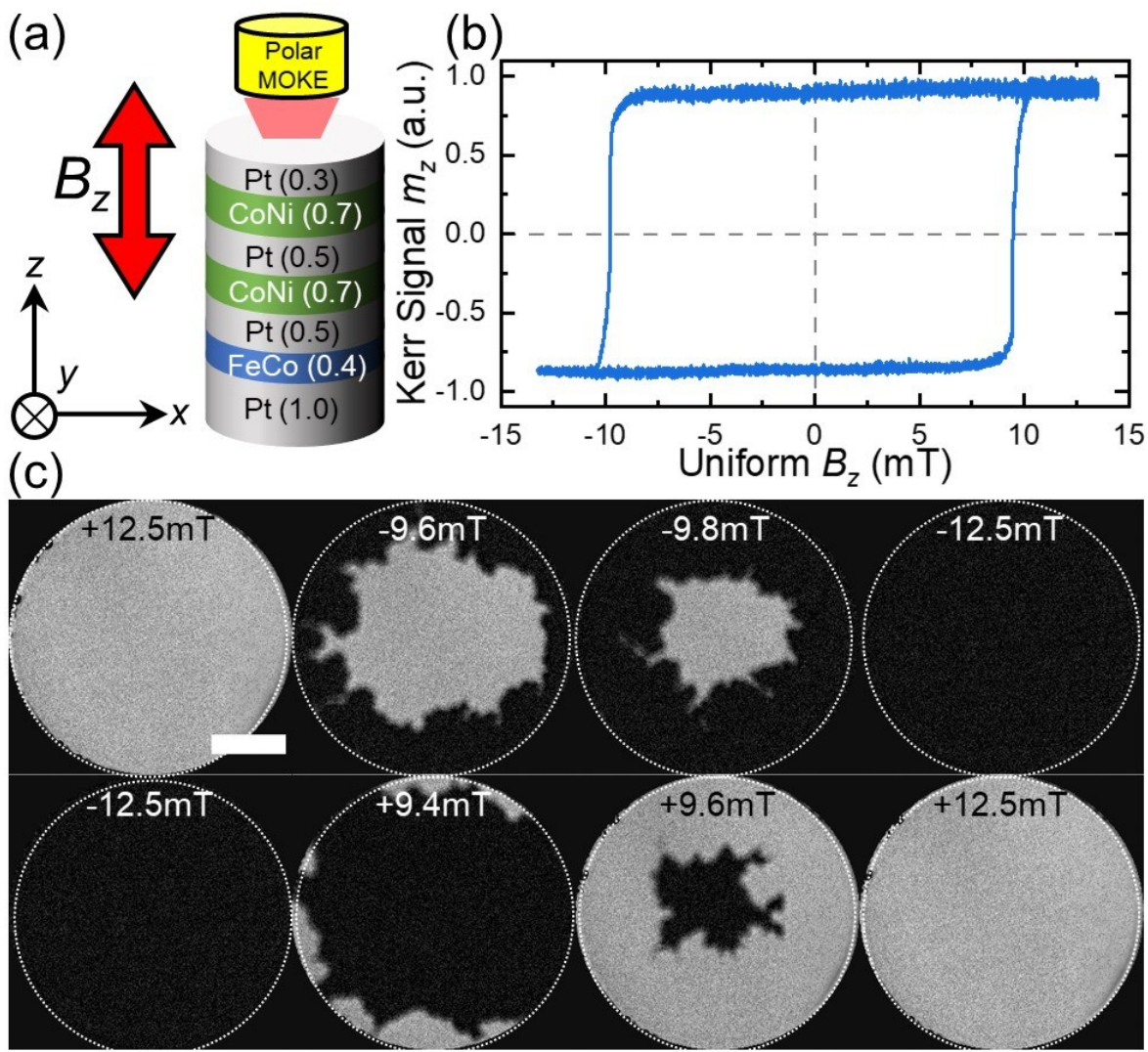

**Figure 5.**

Ohara *et al.*